\documentclass[prd,twocolumn,nofootinbib,showpacs,superscriptaddress]{revtex4}
\usepackage{amsmath}
\usepackage{epsfig}

\newcommand{\eqnref}[1]{Eq.~(\ref{eqn:#1})}
\newcommand{\figref}[1]{Fig.~\ref{fig:#1}}
\newcommand{\secref}[1]{Section~\ref{sec:#1}}
\newcommand{\secsref}[2]{Sections~\ref{sec:#1}-\ref{sec:#2}}

\newcommand{\integral}[4]{\int_{#1}^{#2}{#3}{\mathrm{d}#4}}
\newcommand{\Dh}[0]{D_h}
\newcommand{\tDh}[0]{\tilde D_h}

\begin{document}
\begin{abstract}

Full interpretation of data from gravitational wave observations
will require accurate numerical simulations of source systems, particularly
binary black hole mergers.
A leading approach to improving accuracy in numerical relativity
simulations of black hole systems is through fixed or adaptive mesh
refinement techniques.  We describe a manifestation of numerical interface 
truncation error which
appears as slowly converging, artificial reflections from refinement
boundaries in a broad class of mesh refinement implementations,
potentially compromising the effectiveness of mesh refinement techniques
for some numerical relativity applications if left untreated.  We elucidate this
numerical effect by presenting a model problem which exhibits the
phenomenon, but which is simple enough that its numerical error can be
understood analytically.  Our analysis shows that the effect is caused
by variations in finite differencing error generated across low and high
resolution regions, and that its slow convergence is caused by the
presence of dramatic speed differences among propagation modes typical of 
3+1 relativity.
Lastly, we resolve the problem, presenting a class
of finite differencing stencil modifications, termed mesh-adapted differencing (MAD), 
which eliminate this pathology
in both our model problem and in numerical relativity examples.


\end{abstract}


\title{Reducing reflections from mesh refinement interfaces in numerical relativity}

\author{John G.\ Baker}
\author{James R.\ \surname{van Meter}}
\affiliation{Laboratory for Gravitational Astrophysics, 
NASA Goddard Space Flight
Center, Greenbelt, Maryland 20771}
\maketitle

\section{Introduction}
\label{sec:intro}
Recent years have seen a dramatic rise in opportunities for observing
strong-field gravitational dynamics. New observations of dense
black-hole-like objects, at stellar, intermediate, and supermassive scales are
increasingly frequent.  Anticipated gravitational wave observations by
ground-based and space-based detectors are expected to capture
information about these objects at moments of the strongest
gravitational interactions \cite{Gonzalez:2004nv,Danzmann03}.  Interpretation of data from any such
observations will depend on theoretical modeling of the strong-field
interactions of dense black-hole-like objects in the process of
generating gravitational radiation. General Relativity is the standard
model for describing gravitational interactions and wave generation.
However, the predictions of General Relativity for such cases 
are not yet fully understood, and will depend on 3-D numerical relativity 
computer simulations \cite{Schutz03}.  

While numerical relativity has progressed markedly in recent years \cite{Alcubierre04},
significant improvements in the fidelity of models for events such as
binary black hole coalescence will be essential for the full
interpretation of upcoming observations.  There are many facets to the
problem of improving such simulation, including optimally formulating Einstein's
equations
\cite{Friedrich96,Baumgarte99,Kidder01a,Bona:2003fj}, properly handling boundaries, handling black hole
singularities, handling constraint violations, and making judicious
gauge choices \cite{aei:gauge}.  There are also basic numerical issues concerning how
to, with finite resources, perform such high-fidelity 3-D simulations
with strong short-wavelength gravitational features near the sources,
and weak but critical long-wavelength gravitational (radiation)
features emerging in a large domain.  Approaches to this latter class
of issues include, developing higher order accurate finite
differencing methods \cite{Zlochower:2005bj}, spectral methods \cite{Kidder00a}, mesh refinement techniques \cite{CarpetFMR,Imbiriba04,David} and
other forms of numerical patching techniques
\cite{Calabrese:2004gs,Thornburg:2004dv}.

We focus here on resolving a limitation which has arisen in our work on
numerical relativity simulations of binary black hole systems with
mesh refinement techniques, but which may have analogues in other
numerical patching treatments as well.  Mesh refinement techniques divide the computational domain into regions with separate computational grids which can be of higher resolution in some regions than others. Such approaches involve mesh-structure interfaces across which the details of the finite differencing 
treatment suddenly change.  Inevitably, these interfaces contribute to computational error, manifesting such effect as ``reflections'' off the interfaces.
Considerable attention is given to implementing
``clean'' interfaces, which generate small error, compared to error generated in the bulk regions.  At minimum, this requires interface-induced error to converge at least as rapidly as the bulk error.  For some classes of black hole evolutions, with non-vanishing shift advection terms, we have typically observed 
large error propagating from mesh interfaces.  Understanding and resolving this problem forms the focus of this paper, and we demonstrate a solution, mesh-adapted
differencing (MAD), that works.

\section{Interface Performance}
\label{sec:symptoms}

The pertinent features of our numerical scheme are as follows.
We are solving the 3+1 BSSN formulation of Einstein's equations \cite{Nakamura87,Shibata95,Baumgarte99,Imbiriba04}.  Our gauge condition is numerically determined, typically with some variation of the 1+log slicing and hyperbolic Gamma-driver shift evolution equations \cite{aei:gauge}.  We integrate in time with the iterative Crank-Nicholson method \cite{Teukolsky00}.  All spatial derivatives are computed by second order accurate, centered differencing, save for advection derivatives, for which 
we use second order accurate upwinded differencing.

We are particularly interested in simulating gravitational radiation
generated in black hole collisions.  To resolve the black hole
sources adequately, whilst pushing the computational grid boundary
sufficiently far away, we use fixed mesh refinement (FMR), as
implemented by a software package for this purpose called PARAMESH \cite{MacNeice00}.
With this implementation, the resolutions of two adjacent refinement
regions always differ by a factor of two; i.e. the grid-spacing of the
coarser region is double what it is in the finer region.
``Ghostzones'' or ``guardcells'', typically two layers, are required
to provide buffering between refinement levels.  These guardcells are
filled in by interpolation.

Whether the simulation performs adequately in the presence of
refinement interfaces is a question of particular concern to us.  Inevitably, 
refinement boundaries are sources of numerical errors, although these reflections 
are often satisfactorily convergent and
negligible.  An exception has plagued us in the case of a
non-negligible shift, $\beta^i$, at refinement boundaries.  In this case
we observe a reflection pulse that propagates at a velocity of
$-\beta^i$.  

\begin{figure}
\includegraphics[scale=.36, angle=-90]{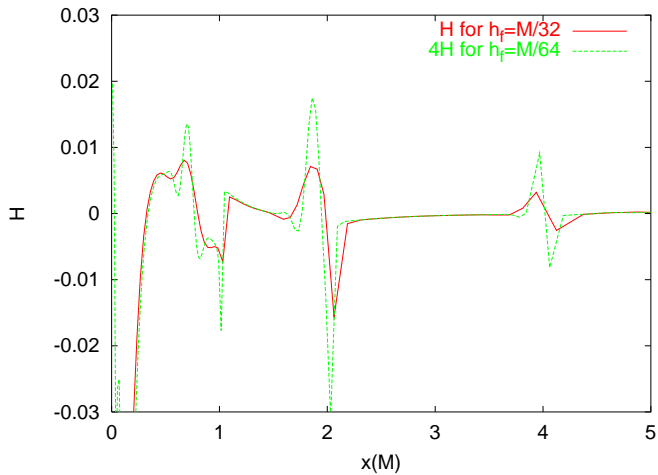}
\caption{Convergence plot for the Hamiltonian constraint error $H$ at time $t=4M$.  There is a single puncture black hole centered at the origin,
and refinement boundaries at $|x_i|=1M$, $2M$, and $4M$.  The finest grid spacing $h_f$ for each simulation is indicated in the figure, and the higher resolution is multiplied by a
factor of 4. For errors demonstrating second order convergence the curves should superpose.} 
\label{BSSN_conv}
\end{figure}

An example can be seen in the case of single
Schwarzschild black hole (in isotropic Schwarzschild coordinates) 
centered in a nested-box
arrangement of refinement regions.  The Hamiltonian constraint
provides a measure of the error in the runs and is plotted in
Fig.~\ref{BSSN_conv}.  
These ``reflected'' error waves, or ``bumps'', notably propagating toward the black hole (at $x=0$) from each interface, are strongest when the
refinement interface is nearest to the black hole (where $\beta^i$ is
largest).  We have also noticed that the reflections seem to originate 
coincidently with the passing of an initial gauge pulse through the
interface.  Such gauge pulses are typical in black hole simulations with ``1+log'' type lapse conditions, and propagate in this case at $\sqrt{2}$ times the speed of light asymptotically.
In Fig.~\ref{BSSN_conv} the convergence of the bumps is indicated by comparing
the error in the Hamiltonian constraint at two resolutions, a moderate resolution run of $h_f=M/32$ and run with uniformly doubled resolution, $h_f=M/64$, where $h_f$ refers to the resolution of the finest grid.  The curves have been rescaled so that they should superpose if the errors are second order convergent. 
Disturbingly, the figure indicates that these errors not to converge
at reasonable grid resolutions, $M/32$ and .  Whether these reflection
errors would converge manifestly at sufficiently high resolution is
difficult to determine in our black hole applications of limited
achievable grid size.  

In any case, the practical effect is that it 
is a challenge to effectively control these errors which may dominate the error
around refinement boundaries in the strong field region.  
Interestingly, however, we have not observed any adverse effect, 
such as poor convergence, imprinted by these bumps on such phenomena as
gravitational radiation measured far from a binary black hole system \cite{David}.
However, other physical quantities of interest, such as the event horizon, 
seem likely to be more sensitive to these near-field errors.

\section{Linearized BSSN}
\label{sec:linearizedbssn}

To understand the source of the $\beta$-speed error exhibited in the last section, we begin by considering a linearized BSSN system with 1+log slicing and hyperbolic Gamma-driver shift.
The system of equations is:
\begin{eqnarray}
{\dot a} &=& -2K\\
{\dot B^i} &=& {\dot{\tilde \Gamma^i}}\\
{\dot \beta_1^i} &=& \frac{3}{4} B\\
{\dot \phi} &=& -{1\over 6}(K - \partial_i \beta_1^i)+\beta_0^k\partial_k\phi\\
{\dot K} &=& - \partial_i \partial_i a +\beta_0^k\partial_k K\\
{\dot h}_{ij} &=& -2 {\tilde A}_{ij} + \partial_i \beta_1^j + \partial_j\beta_1^i - {2\over 3}\delta_{ij} \partial_k\beta_1^k +\beta_0^k\partial_k h_{ij}\\
{\dot{\tilde A}}_{ij} &=& \left[ -\partial_i\partial_ja - {1\over 2}\partial_k\partial_kh_{ij} 
+ {1\over 2} \partial_i{\tilde\Gamma}^j + {1\over 2} \partial_j{\tilde\Gamma}^i\right. \nonumber \\ && \left.- 2\partial_i 
\partial_j\phi \right]^{\rm TF}
  +\beta_0^k\partial_k {\tilde A}_{ij}\\
\dot{\tilde\Gamma}^i &=& -{4\over 3} \partial_i K + \partial_k\partial_k \beta_1^i + {1\over 3} \partial_i \partial_j \beta_1^j  +\beta_0^k\partial_k {\tilde\Gamma}^i
\end{eqnarray}
where $a \equiv \alpha-1$, $\beta_1^i \equiv \beta^i-\beta_0^i$, (with $\beta_0^i$ assumed spatially uniform for simplicity), and $h_{ij} \equiv
{\tilde \gamma_{ij}}-\delta_{ij}$.

For a problem that varies only in one dimension, with no initial transverse components, if we assume plane-wave solutions (generalizable by Fourier analysis), then the above system of equations can be written in the form:
\begin{eqnarray}
\label{ket_evolution}
\partial_t|u\rangle = ik\mathsf{M}|u\rangle 
\end{eqnarray}
where
\begin{equation}
|u\rangle = \left( \begin{array}{c} {\hat a} \\ {\hat B} \\ {\hat \beta_1} \\ {\hat \phi} \\ {\hat K} \\ {\hat h} \\ {\hat A} \\ {\hat \Gamma}   \end{array}  \right)e^{i(kx-\omega t)} 
\end{equation}
with ${\hat a}$, ${\hat B}$, ${\hat \beta_1}$, ${\hat \phi}$, ${\hat K}$, ${\hat h}$, ${\hat A}$, and ${\hat \Gamma}$ the amplitudes of $a$, $B^x$, $\beta_1^x$, $\phi$, $K$, $h_{xx}$, ${\tilde A}_{xx}$, and ${\tilde \Gamma}^x$ respectively,
and 
\begin{equation}
 \mathsf{M}= \left( \begin{array}{c c c c c c c c} 
       0         &        0        &         0       &       0       &  \frac{2}{ik} &       0       &       0       &       0       \\
       0         &        0        &   \frac{4ik}{3} &       0       &  \frac{4}{3}  &       0       &       0       &-\beta_0       \\
       0         & \frac{3}{4ik}   &         0       &       0       &       0       &       0       &       0       &       0       \\
       0         &        0        &  -\frac{1}{6}   &  -\beta_0     & -\frac{1}{6ik}&       0       &       0       &       0       \\
       -ik       &        0        &         0       &       0       &  -\beta_0     &       0       &       0       &       0       \\
       0         &        0        &   -\frac{4}{3}  &       0       &        0      &   -\beta_0    & -\frac{2}{ik} &       0       \\
 -\frac{2ik}{3}  &        0        &         0       & -\frac{4ik}{3}&        0      & -\frac{ik}{2} &  -\beta_0     & -\frac{2}{3}  \\
       0         &        0        & \frac{4ik}{3}   &       0       &  \frac{4}{3}  &       0       &       0       &   -\beta_0    
\end{array} \right)            
\end{equation}

The eigenvalues are
$0$,$-\beta_0$,$-\beta_0+1$,$-\beta_0-1$,$-\frac{1}{2}\beta_0+\frac{1}{2}\sqrt{\beta_0^2+8}$,$-\frac{1}{2}\beta_0-\frac{1}{2}\sqrt{\beta_0^2+8}$,$-\frac{1}{2}\beta_0+\frac{1}{2}\sqrt{\beta_0^2+4}$, and $-\frac{1}{2}\beta_0-\frac{1}{2}\sqrt{\beta_0^2+4}$.
If $|\lambda \rangle$ is the eigenvector associated with eigenvalue $\lambda$, and $\langle \lambda |$ is defined such that $\sum_{\lambda} |\lambda\rangle\langle \lambda| = \mathsf{I}$, 
then 
\begin{equation}
\label{ket_operator}
\mathsf{M}=\sum_{\lambda} \lambda |\lambda\rangle \langle \lambda|  
\end{equation}
By substituting Eq.~(\ref{ket_operator}) into Eq.~(\ref{ket_evolution}) the system of evolution equations can be decomposed into a series of advection terms, each associated with a characteristic
velocity equal to one of the eigenvalues.  Note that, assuming $\beta_0 \ll 1$, the $\beta$-speed mode is much slower than any of the other non-zero-speed modes.  As we are particularly concerned with this mode, it is instructive to write the evolution
equation thus:

\begin{eqnarray}
\label{eqn:linBSSNadvectionResult}
 \partial_t  \left( \begin{array}{c} { a} \\ { B} \\ { \beta_1} \\ { \phi} \\ { K} \\ { h} \\ { A} \\ { \Gamma}   \end{array} \right)&=&-\beta_0 \partial_x  \left( \begin{array}{r} 0 \\ 0 \\ 0  \\ -\frac{3}{8} \\ 0 \\ 1 \\ 0 \\ 0   \end{array} \right) \left( -\frac{8}{3}\phi  -\frac{1}{3ik}\Gamma  \right) \nonumber \\ &&+\sum_{\lambda\neq-\beta_0} \lambda \partial_x|\lambda\rangle \langle \lambda |u\rangle
\end{eqnarray}

This equation indicates that disturbances originating in $\phi$ and ${\tilde \Gamma}^i$ can propagate in $\phi$ and $h_{ij}$ at $\beta$-speed and, further, that this is the only means of
$\beta$-speed propagation allowed by this system.  Thus $\phi$ is uniquely significant in both generating and propagating $\beta$-speed modes.
Note that \eqnref{linBSSNadvectionResult} resembles the $\dot \phi$ evolution equation in the BSSN system.

\section{A model problem}
\label{sec:bumpy}

Motivated by the discussion of the last section, 
a simple model for the generation and propagation of the $\beta$-speed
modes is a one-dimensional advection
problem with an additional driving term,
\begin{equation}
\label{eqn:bumpy}
\dot \varphi(x,t)=\beta \partial_x\varphi(x,t)+f(x,t),
\end{equation}
In our numerical simulations, it appears that the reflected ``bumps'' may be
triggered by a rapidly propagating gauge pulse which propagates
outward early in the simulations.  For our model problem, which we will term
``Bumpy'' because of its most salient feature, we will drive
the advection equation with a pulse propagating at speed $v$,
significantly faster than the advection speed $\beta$. We thus let
$f(x,t)=f(x-vt)$ where $v$ is larger than $\beta$ and both are assumed
positive.  This model equation is simple enough
to be solved directly,
\begin{eqnarray}
\varphi(x,t)&=&\integral{0}{t}{f(x+(t-t')\beta,t')}{t'}\nonumber\\
&=&\frac{-1}{v+\beta}(F(x-vt)-F(x+\beta t))\label{eqn:bumpy-solve}
,
\end{eqnarray}
where $\partial_x F(x)=f(x)$. 
For definiteness we can take the driving term to be a Gaussian pulse,
$f(x)=\exp(-x^2)$, implying $F(x)=(\sqrt(\pi)/2)\mbox{erf}(x)$.  For a
localized pulse, such as this, the $F(x+\beta t)$ term in the exact
solution will be negligible in the region of interest, near $x=x_0$.

\begin{figure}
\includegraphics[scale=.3, angle=-90]{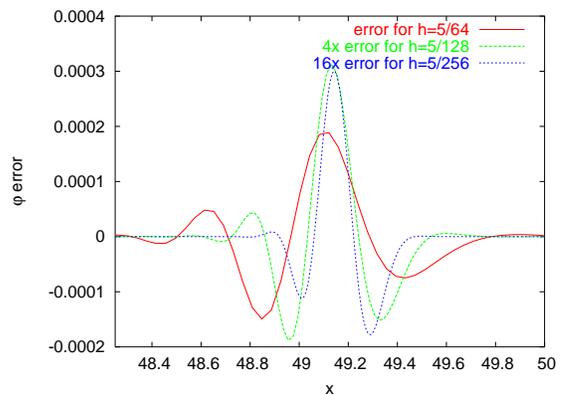}
\caption{Convergence plot for the error in $\varphi$.  There is a refinement boundary at $x=50$. The coarse grid spacing $h$ of each simulation is indicated in the
figure, and the medium and high resolution curves have been multiplied by $4$ and $16$ respectively, as appropriate to demonstrate second order convergence. }
\label{fig:BUMP_conv}
\end{figure}

To test if this model problem is sufficient, we have numerically
evolved \eqnref{bumpy} using a 1-D finite difference code with a
resolution $dx=h$ for $x<x_0$ and $dx=2h$ for $x>x_0$, realizing a
refinement boundary at $x=x_0=50$.  We evolved over the domain $0<x<100$
with periodic boundary conditions, using three-point upwind finite 
difference stencil and a mesh refinement scheme
similar to that in our numerical simulations with PARAMESH.  
In \figref{BUMP_conv} we show the errors in our evolution variable
$\varphi$ in the vicinity of the refinement interface
for runs with several resolutions, $h=5/64, 5/128$ and $5/256$ 
.  In each case there clearly a reflection 
error bump propagating to the left, away from the refinement interface,
which we find propagates at speed $\beta$.
As before the errors have been rescaled so that
that second order converging error should superpose. 
Comparing the two lower resolution run, we do not see good superposition, 
and the peaks appear to converge at roughly first order.  The comparative appearance of these errors is similar to that seen for black hole simulations in
Fig.~\ref{BSSN_conv}. 
However, running at higher resolutions, which is readily allowed by this 1-D model, we
find that the reflected error is indeed second order convergent, as is suggested by comparing the two higher resolution runs in  \figref{BUMP_conv}.
This convergence is manifest only at relatively high resolution
(high relative to the wavelength -- and higher than can be easily achieved
with respect to a gravitational wavelength in a binary black hole run).  
Further experimentation has indicated that these results are not strongly 
affected by variations in the 
time-integration method or by changing among mesh refinement interfacing 
schemes which are consistent with the overall second order finite differencing 
accuracy.

\section{A finite difference analysis}
\label{sec:bumpyfd}

In this section we attempt to understand the numerical behavior of the
Bumpy problem \eqnref{bumpy} by constructing here an analytic model
for the numerical error in our simulations.  Noting that the time 
discretization, and the details of the interpolation scheme used in applying 
refinement conditions seem not to be directly linked to the problematic 
error features we see, we model the error with as few assumptions as 
possible about these details.  We treat the numerical errors 
continuously in time, and we consider the effects of spatial finite 
differencing in terms of a continuous field $\varphi(x,t,h)$ representing the
numerical solution at (fine-grid) resolution $dx=h$.  We then expand 
$\varphi$ in orders of $h$,
\begin{equation}
\varphi(x,t,h)=\varphi_e(x,t)+h^2\varphi_2(x,t)+{\cal O}(h^3),
\end{equation}
Where $\varphi_e(x,t)$ is understood to be the exact solution 
\eqnref{bumpy-solve}.
We consider the effect of our finite differencing scheme by replacing
the spatial derivative appearing in \eqnref{bumpy} with a suitable
finite difference operator $\tDh$. Thus,
\begin{equation}
\label{eqn:bumpyfd-start}
\dot \varphi(x,t,h)=\beta \tDh\varphi(x,t,h)+f(x,t).
\end{equation}
As above, we include a refinement jump in the grid at $x=x_0$, 
represented here by
applying a coarser version of the finite difference stencil in the
$x>x_0$ part of the spatial domain.  This will be consistent with any
refinement interface algorithm which applies the same finite
difference stencil in both the coarse and fine regions, and applies a
interpolative guard-cell filling algorithm at the interfaces which
leads to finite differences at the interface which are consistently
second order accurate.  On a uniform grid the second order error term of
a finite first derivative is generally proportional to the third
derivative of the field,
\begin{equation*}
\Dh\varphi(x,t,h)=\partial_x\varphi(x,t,h)+e_2
h^2\partial_{x}^{3}{\varphi(x,t,h)}+{\cal O}(h^3).
\end{equation*}
For the specific upwind differencing operator used in \secref{bumpy}, the stencil of
which is, 
\begin{eqnarray}
  D_{h} &=& \frac{1}{h} \left( -\frac{3}{2} + 2E_h - \frac{1}{2}E_{2h} \right)
\end{eqnarray}
where $E$ is the spatial translation operator defined such that $E_hf(x) = f(x+h)$, 
the constant error coefficient turns out to be
$e_2=-1/3$.
Including the refinement jump, we have
\begin{eqnarray}
\tDh\varphi&=&\Dh\varphi+\Theta(x-x_0)(D_{2h}-\Dh)\varphi\label{eqn:Dh}\\
&=&\partial_x\varphi+(1+3\Theta(x-x_0))e_2h^2\partial_{x}^{3}{\varphi}+{\cal O}(h^3).\nonumber
\end{eqnarray}
Substituting into \eqnref{bumpyfd-start}, and rearranging, yields
\begin{eqnarray}
h^2\dot\varphi_2(x,t,h)&=&\left[-\dot\varphi_e(x,t)+\beta \partial_x\varphi_e(x,t)+f(x,t)\right]\nonumber\\
&&+h^2\beta \big[\partial_x\varphi_2(x,t,h)+\nonumber\\
&&\qquad e_2(1+3\Theta(x-x_0))\partial_{x}^{3}{\varphi_e(x,t)}\big]\nonumber\\
&&+{\cal O}(h^3).
\end{eqnarray}
Then noting that the first term vanishes, and taking the limit $h\rightarrow0$, we derive
\begin{eqnarray}
\dot\varphi_2(x,t)&=&\beta\partial_x\varphi_2(x,t)+\label{eqn:bumpyfd}\\
&&\qquad\beta e_2(1+3\Theta(x-x_0))\partial_{x}^{3}{\varphi_e(x,t)}\nonumber,
\end{eqnarray}
where we have used the notation $\varphi_2(x,t)=\varphi_2(x,t,0)$.
This is our model for the generation and propagation of finite differencing error in the numerical model problem in \secref{bumpy}.

Since \eqnref{bumpyfd} is of the same form as \eqnref{bumpyfd} we can  
solve it the like fashion.  
We leave out the negligible $F(x+\beta t)$ in $\varphi_e$, 
substituting
\begin{eqnarray*} 
f(x,t)&=&\beta e_2(1+3\Theta(x-x_0))\partial_{x}^{3}{\varphi_e(x,t)}\\
&\simeq&-\frac{\beta e_2}{v+\beta}(1+3\Theta(x-x_0))
\partial_{x}^{3}{F(x-vt)}
\end{eqnarray*}
into the first line of \eqnref{bumpy-solve}. Then, performing the
integral with careful attention to the presence of the step function,
we get the solution
\begin{eqnarray}
\varphi_2(x,t)&=&(s(x-vt)-s(x+\beta t))(1+3\Theta(x-x_0))\nonumber\\
&&+3s\big(\frac{-v}\beta(x-x_0+\beta(t-\frac{x_0}v)\big)\nonumber\times\\
&&\qquad (\Theta(x-x_0+\beta t)-\Theta(x-x_0)) \label{eqn:bumpyfd-solve}
\end{eqnarray}
where,
\begin{eqnarray*}
s(x)&=&\frac{\beta e_2}{(v+\beta)^2}\partial_x^{2}F(x)\\
 &=&-\frac{2 \beta e_2 x}{(v+\beta)^2}\exp{(-x^2)}.
\end{eqnarray*}
As in \secref{bumpy}, the $s(x+\beta t)$ is negligible in the relevant region 
near the refinement interface. Thus the first term in
\eqnref{bumpyfd-solve} is effectively the differencing error
associated with the upsweep in $\varphi_e$ which propagates across the
grid at speed $v$.  This part grows four times larger in the coarse
$x>x_0$ region.  The second term is quite interesting, as it propagates
in the reverse direction at speed $\beta$.  It has the same $s(x)$
shape as the forward propagating component, but it is reversed, and
contracted by a factor $\beta/v$.  It is timed to originate at the
interface as the first pulse crosses.  The two terms combine make the
full solution continuous at the interface.  Heuristically, one could
say that the second term is caused by the discontinuity in the
differencing error, generated in order to produce a regular solution,
and then, necessarily advecting away as required by the original model
equation \eqnref{bumpy}.  It is contracted because it propagates at a 
different speed than the first term, but their time dependences must match 
at the interface.

Note that although we concretely consider second order finite
differencing here, the finite differencing order is largely irrelevant
in this analysis.  The calculation can be directly adapted 
for the leading order in a higher order finite differencing 
by changing a few coefficients.

\begin{figure}
\includegraphics[scale=.3, angle=-90]{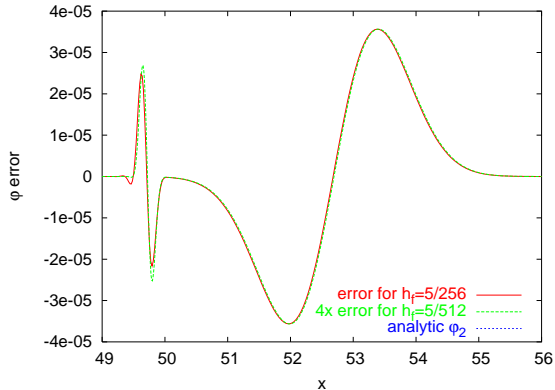}
\caption{Numerical error in $\varphi$, immediately after a pulse
incident on the refinement interface at $x=50$ has passed through,
generating a pulse of transmitted error and a contracted pulse of
reflected error.  The error is second order convergent and in
agreement with the analytic prediction of the numerical error, as
indicated.}  
\label{fig:BUMP_annu}
\end{figure}
We check that this analysis accurately describes the numerical errors in our 
Bumpy model evolutions by comparing the predicted error with the numerical 
error results.  As Fig.~\ref{fig:BUMP_annu} shows,
the prediction of \eqnref{bumpyfd-solve} agrees to high precision with the
numerical simulation errors realized in high resolution runs.
The figure shows both the rightward propagating wave which travels at velocity
$v$ with the driving pulse, and the shorter-wavelength reflected wave.
The reason for the slow convergence in our Bumpy simulations is now clear.
Because the reflection must propagate at a significantly slower speed than the
pulse which generated it, while their frequencies must match, the reflected 
pulse necessarily has a shortened wavelength.  The key ingredients in this 
producing this effect are that we are simulating a system with 
strongly mismatched propagation speeds, and with the potential for generating 
mode-mixing reflections off numerical features such as our refinement 
interfaces such.  In such a situation, with resolution
just sufficient to accurately resolve the long-wavelength features in the 
solution, the shortened wavelength of the numerical reflections makes
them unresolvable without a significant increase in the resolution 
(say by a factor of $v/\beta$).  We thus have reason to expect that such an 
effect is indeed the source of the bumpy reflection in our black hole runs.
We will verify this by attempting to eliminating this type of error.

\section{Improving the differencing stencils}
\label{sec:thefix}

Slowly converging errors of the type exposed in \secsref{bumpy}{bumpyfd} 
now seem likely to occur under fairly general circumstances.  This 
explains our difficulties in many attempts to eliminate these 
effects by tweaking the
interface conditions in various ways.  Our analysis suggests that the only
ways to avoid these problems, aside from eliminating the slowly propagating 
modes by setting $\beta^i=0$, is to remove the mode-mixing reflections at 
interfaces.
The results of the last section clearly show that the reflected error in the
Bumpy system is overwhelmingly due to the discontinuity in the differencing 
operator.  More specifically, the reflection is related to the
discontinuity in the second order truncation error.  
This observation suggests a solution.

Consider using modified differencing stencils such that the coefficient
of the second order truncation error in the fine grid region is multiplied by a
constant factor $q_0$ while the coefficient of the second order truncation error in
the coarse grid region is multiplied by a constant factor $q_1$.  Then
the previously constant coefficient $e_2$ of the last section becomes:

\begin{eqnarray}
e_2 \to [q_0+(q_1-q_0)\Theta(x-x_0)]e_2
\end{eqnarray}





Repeating the steps of the last section for this new truncation error, 
%
%
the solution for the second order error in $\varphi$ becomes,
\begin{eqnarray}
\varphi_2(x,t)&=&(s(x-vt)-s(x+\beta t))(q_0+(4q_1-q_0)\Theta(x-x_0))\nonumber\\
&&+(4q_1-q_0)s\big(\frac{-v}\beta(x-x_0+\beta(t-\frac{x_0}v)\big)\nonumber\times\\
&&\qquad(\Theta(x-x_0+\beta t)-\Theta(x-x_0)) \label{eqn:bumpilessfd-solve}
\end{eqnarray}
Thus for the spatially blueshifted, reflected error, we now obtain,
\begin{eqnarray}
\varphi_{ref}=(4q_1-q_0)s(-\frac{v}{\beta}(x-x_0+\beta(t-\frac{x_0}{v})))  
\end{eqnarray}

One possibility would be to eliminate the leading order reflection
error simply by choosing $q_0=q_1=0$.  This choice corresponds to
using higher order stencils, say third or fourth order accurate,
throughout the entire grid.  
Higher order differencing methods are clearly valuable in numerical
relativity simulations \cite{Zlochower:2005bj}.
However, we will focus here on identifying a minimal way to realize 
effective second order convergence in a second order convergent finite 
differencing scheme.  Our result should generalize to arbitrary order.

Note that with the choice $q_0=1$ and $q_1=\frac{1}{4}$, the reflected error vanishes.
More generally, of course, any choice of $q_0$ and $q_1$ that makes the truncation error continuous
across the refinement boundary will remove the second order reflection.  In particular, the choice
\begin{eqnarray}
q_n = \left(\frac{h_0}{h_n}\right)^2
\end{eqnarray}
will work in the $n$th refinement region of a grid with an arbitrary number of refinement levels, where $h_0$ is assumed to be the 
grid-spacing in the finest region.  We call such a mesh-adapted differencing scheme MAD.

A second order MAD stencil can be obtained simply by linearly combining a second order accurate stencil with
a higher order accurate stencil, as follows:
\begin{eqnarray}
\label{MAD_general}
D_{h_n} = q_nD_{h_n}^{[2]} + (1-q_n)D_{h_n}^{[j]}
\end{eqnarray}
where the superscripted numbers in square brackets represent the order of accuracy of the differencing operator, and $j>2$.  (Of course, this stratagem can be readily generalized for higher order MAD operators as well.)  In the particular case of a second order 
upwinded stencil combined with a third order lopsided stencil, the resulting stencil has the form:
\begin{eqnarray}
\label{MAD_upwind}
  D_{h_n} &=& \frac{1}{h_n} \left[ \left( -\frac{1}{3}+\frac{1}{3}q_n \right) E_{-h_n} + \left( -\frac{1}{2}-q_n \right) \right. \nonumber \\ && \left.  + \left( 1+q_n
  \right)E_{h_n} + \left(
-\frac{1}{6}-\frac{1}{3}q_n \right)E_{2h_n} \right] 
\end{eqnarray} 

\begin{figure}
\includegraphics[scale=.3, angle=-90]{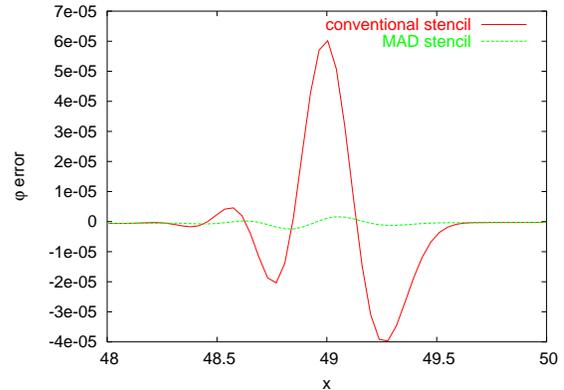}
\caption{Comparison of the error in $\varphi$ reflected from the refinement boundary at $x=50$ in the case of a second order conventional stencil and a second order MAD stencil.}
\label{BUMP_MAD}
\end{figure}
We have implemented this MAD stencil in the numerical simulations of the 
Bumpy system. Fig.~\ref{BUMP_MAD} shows the result for a run at a moderate 
resolution.  With the MAD stencil
the dominant reflected error is nearly eliminated.




\section{Results for black hole evolutions}
\label{sec:results}

We can demonstrate that our analysis does indeed explain the problematic reflections in our black hole simulations, by verifying that the same remedies applied to fix our Bumpy model problem, i.e. higher order or MAD stenciling, also 
fix the problem in our black hole simulations.
 
As in \cite{Zlochower:2005bj}, we have found that third or higher order accurate differencing
of advection terms is unstable in the vicinity of black hole punctures, if the stencil has any points on
the ``downwind'' side.  If the 3rd or higher order accurate differencing stencil does not have
any points on the downwind side, it requires three or more layers of guardcells to accommodate
points on the upwind side, which is expensive memory-wise.  
Thus, although higher order spatial differencing should reduce reflection
from refinement boundaries, it brings with it a new set of problems to solve.

Use of a second order accurate MAD stencil, as in Eq.~(\ref{MAD_upwind}), for advection, avoids the above difficulties.  As we generally
locate the punctures within the finest grid regions, and the MAD stencil automatically reverts to conventional second order
upwinding in this region, the advection derivative does not take any points on the downwind side in 
the vicinity of the puncture and we find that stability is maintained.

In our Einstein solver, we have implemented a second order accurate
MAD stencil for advection.  For all non-advection derivatives, the MAD
stencil is constructed from a linear combination of second order
centered and fourth order centered stencils, as in
Eq.~(\ref{MAD_general}).  As a result, reflection from refinement
boundaries has been dramatically reduced in the case of a single
puncture. 
\begin{figure}
\includegraphics[scale=.3, angle=-90]{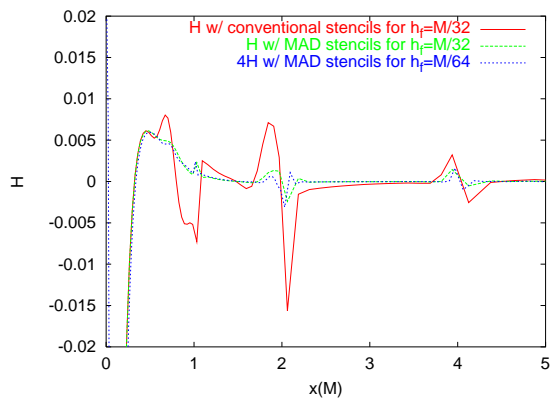}
\caption{Comparison of the Hamiltonian constraint $H$ in the case of second order accurate, conventional differencing stencils
and second order accurate, MAD stencils, as well as a demonstration of convergence in the latter case.   There is a single puncture black hole centered at the origin,
and refinement boundaries at $|x_i|=1M$, $2M$, and $4M$.}
\label{BSSN_MAD}
\end{figure}
Fig.~\ref{BSSN_MAD} compares the Hamiltonian
 constraint error from a run with conventional stencils with the Hamiltonian
 constraint error in runs with MAD stencils, demonstrating clear improvement.
The plot shows that the reflection error for our $h_f=M/32$ case is
much reduced with the MAD stencil.  By comparing with the $h_f=M/64$, 
MAD-stencil run, we also see that the remaining bump now converges away at 
second order or better.

For simplicity we only show plots from single black hole simulations here, 
though we have applied MAD in more interesting binary black hole cases as well.
We find that the improvements generalize naturally to these cases.
In particular we find that errors in the
Weyl scalars near refinement boundaries are reduced by at least an
order of magnitude.

\section{Conclusions}
\label{sec:conclusions}

We have studied a problem which occurs in numerical relativity simulations 
with nonvanishing shift with mesh refinement. These involve slow advection 
modes across a mesh interface, and produce slowly converging numerical 
reflections that propagate at the speed $\beta$ of the advection.
We have successfully modeled the problem analytically.  Our analysis suggests 
that the the effect is a general consequence of discontinuous jumps in the 
finite differencing stencil in a problem with propagation modes of widely differing speeds.  Our proposed solution, to adjust the finite differencing 
stencils to make the leading order differencing error 
continuous across mesh interfaces, is shown to be effective in black hole simulations with mesh refinement, but may have wider application in numerical relativity.
In addition to grids of non-uniform refinement, mesh-adapted differencing may also be appropriate for grids with multiple coordinate patches,
where discontinuous differencing error can also be expected.

\begin{acknowledgments}
We are happy to thank J. David Brown for helpful discussion.
We gratefully acknowledge CPU time grants from the Commodity
Cluster Computing Project (NASA-GSFC) and Project Columbia (NASA
Advanced Supercomputing Division, NASA-Ames).  This work was supported
in part by NASA Space Sciences grant ATP02-0043-0056. JvM was also
supported in part by the Research Associateship Programs Office of the
National Research Council.
\end{acknowledgments}

\bibliography{phren}

\end{document}